\documentclass[aps,prd,reprint,floatfix,nofootinbib,superscriptaddress]{revtex4-2}

\usepackage[T1]{fontenc}
\usepackage[utf8]{inputenc}
\usepackage{amsmath,amssymb,bm}
\usepackage{graphicx}
\usepackage{booktabs}
\usepackage{array}
\usepackage{siunitx}
\usepackage{xcolor}
\usepackage{microtype}
\usepackage{hyperref}
\hypersetup{colorlinks=true,linkcolor=blue,citecolor=blue,urlcolor=blue}

\newcommand{\mchi}{m_{\widetilde{\chi}_1^0}}
\newcommand{\mchiTwo}{m_{\widetilde{\chi}_2^0}}
\newcommand{\mchip}{m_{\widetilde{\chi}_1^\pm}}
\newcommand{\dmchi}{\Delta m_{21}}
\newcommand{\Omegachi}{\Omega_{\chi}h^2}

\begin{document}

\title{The compressed bino--wino dark matter in the GmSUGRA and the LHC soft-lepton searches}

\author{Imtiaz Khan}
\email{ikhanphys1993@gmail.com}
\affiliation{Department of Physics, Zhejiang Normal University, Jinhua, Zhejiang 321004, China}
\affiliation{Research Center of Astrophysics and Cosmology, Khazar University, Baku, AZ1096, 41 Mehseti Street, Azerbaijan}
\author{Ali Muhammad}
\email{alimuhammad@phys.qau.edu.pk}
\affiliation{CAS Key Laboratory of Theoretical Physics, Institute of Theoretical Physics, Chinese Academy of Sciences, Beijing 100190, China}
\affiliation{School of Physical Sciences, University of Chinese Academy of Sciences, No. 19A Yuquan Road, Beijing 100049, China}

\author{Tianjun Li}
\email{tli@itp.ac.cn}

\affiliation{School of Physics, Henan Normal University, Xinxiang 453007, P. R. China}

\author{Shabbar Raza}
\email{shabbar.raza@fuuast.edu.pk}
\affiliation{Department of Physics, Federal Urdu University of Arts, Science and Technology, Karachi 75300, Pakistan}

\author{Mussawir Khan} 
\email{mussawirkhan@ihep.ac.cn}
\affiliation{State Key Laboratory of Particle Astrophysics, Institute of High Energy Physics, Chinese Academy of Sciences, Beijing 100049, China}
\affiliation{University of Chinese Academy of Sciences, Beijing 100049, China}

\begin{abstract}
The search for electroweak supersymmetric particles remains an important part of the LHC program since neutralinos and charginos with masses around a few hundred GeV are still compatible with the present constraints in compressed spectra. Run-2 searches for the ``golden channel'', $pp\to\widetilde\chi_2^0\widetilde\chi_1^\pm\to\widetilde\chi_1^0 Z^{(*)}\widetilde\chi_1^0 W^{\pm(*)}$, have reported mild excesses in parts of the soft two- and three-lepton data, with the corresponding simplified-model interpretations emphasizing $m_{\widetilde\chi_2^0}\simeq m_{\widetilde\chi_1^\pm}\gtrsim200$~GeV and a neutralino mass difference near $20$~GeV. We study this configuration in the Minimal Supersymmetric Standard Model (MSSM) with the Generalized minimal Supergravity (GmSUGRA) boundary conditions, imposing the Higgs, flavor, collider, relic-density and direct-detection constraints relevant to the spectrum. For both signs of $\mu$, the solutions compatible with the Planck experiment contain an almost pure bino $\widetilde\chi_1^0$ with mass about $260$--$266$~GeV and a nearly degenerate wino-like $\widetilde\chi_2^0/\widetilde\chi_1^\pm$ pair near $280$~GeV. The resulting splitting, $\Delta m_{21}\simeq18.5$--$19.1$~GeV, lies in the mass-difference region probed by the ATLAS and CMS soft-lepton searches and places the $Z^{(*)}/W^{(*)}$ decays below the on-shell gauge-boson thresholds. The same compressed hierarchy keeps the wino states thermally populated during freeze-out, so bino-wino coannihilation yields a neutralino relic abundance consistent with the Planck experiment while the surviving solutions remain compatible with the direct-detection experimental bounds. Thus, the GmSUGRA boundary conditions provide a high-scale realization of the compressed bino-wino spectrum relevant to the LHC soft-lepton channel.
\end{abstract}

\maketitle

\section{Introduction}
\label{sec:intro}

Supersymmetric extensions of the Standard Model (SM) provide a neutral dark-matter candidate when $R$ parity is conserved and allow the electroweak and colored superpartner masses to be organized within a common high-scale framework~\cite{Dimopoulos:1981yj,Ellis:1990wk,Amaldi:1991cn,Goldberg:1983nd,Ellis:1983ew,Jungman:1995df,Feng:2010gw}. LHC searches have pushed large parts of the strongly produced squark and gluino spectrum to the multi-TeV scale, while neutralinos, charginos, and sleptons can remain substantially lighter~\cite{ATLAS:2020dsf,ATLAS:2019gdh,Cheng:2012np,Khan:2026ewsusy}. This separation leaves electroweakino production as a direct probe of supersymmetry at the electroweak scale.

A challenging configuration is obtained when the lightest neutralino is followed by a nearly degenerate second neutralino and chargino. The process
\begin{equation}
pp\to\widetilde\chi_2^0\widetilde\chi_1^\pm\to\widetilde\chi_1^0 Z^{(*)}\widetilde\chi_1^0 W^{\pm(*)}
\label{eq:production}
\end{equation}
produces missing transverse momentum together with leptons whose momenta are set by the available mass difference. For $\dmchi\equiv\mchiTwo-\mchi<m_Z$, the neutralino decay is three-body, and, neglecting the charged-lepton masses, the dilepton invariant mass satisfies approximately $m_{\ell\ell}^{\rm max}\simeq\dmchi$.
The sensitivity therefore depends on low-$p_T$ lepton reconstruction, the missing-momentum requirement and recoil against initial-state radiation.

ATLAS and CMS have developed two- and three-lepton searches for this compressed regime~\cite{ATLAS:2020compressed,ATLAS:2021threelep,CMS:2022soft}. The CMS Run-2 soft-lepton analysis excludes wino-bino masses up to about $275$~GeV for a $10$~GeV splitting in the corresponding simplified model~\cite{CMS:2022soft}. The later CMS and ATLAS Run-2 combinations broaden the coverage of electroweakino parameter space and improve the sensitivity relative to individual searches~\cite{CMS:2024combined,ATLAS:2024combination}. CMS has also extended the reconstruction of very soft electrons and isolated tracks, improving the reach for mass differences of a few GeV or below~\cite{CMS:2025softupdate,CMS:2025leptrack,CMS:2026track}. These analyses probe a region distinct from conventional on-shell $WZ$ searches, where the visible decay products are substantially harder.

Several phenomenological studies have examined the mild upward fluctuations reported in parts of the compressed-spectrum data. A combined discussion of ATLAS and CMS soft-lepton searches associates the relevant dilepton masses mainly with the $10$--$20$~GeV interval~\cite{Agin:2024coherent}. Within the MSSM, wino--bino spectra with a few-hundred-GeV wino and a bino lighter by roughly $20$~GeV reproduce the corresponding kinematics~\cite{Chakraborti:2024pdn}. Recasts combining soft-lepton and monojet searches select a related wino--bino region~\cite{Agin:2025joint}, while independent MadAnalysis~5 implementations provide a framework for testing the compressed ATLAS signal regions beyond the experimental simplified models~\cite{Araz:2026ma5}. A recent global electroweakino analysis, which combines a substantially larger set of Run-2 searches and measurements, gives more restrictive bounds within its decoupled-sfermion EWMSSM setup and does not obtain a common fit to all compressed-spectrum fluctuations~\cite{Athron:2026ewino}. The simplified-model overlays and the global result therefore address different assumptions and are both relevant when a complete supersymmetric spectrum is considered.

The relic abundance supplies an independent relation between the same masses. A bino-dominated LSP has suppressed annihilation when the other superpartners are well separated in mass. If a wino-like neutralino and chargino lie nearby, their electroweak interactions contribute through coannihilation and can reduce the thermal bino abundance~\cite{Griest:1990kh,Edsjo:1997bg}. The mass difference that governs the soft-lepton kinematics at the LHC thus also controls the thermal population of the wino states. Additional light sleptons can further modify both freeze-out and the collider decay pattern.

Generalized minimal supergravity (GmSUGRA) provides a high-scale setting in which such a spectrum can arise from nonuniversal but correlated gaugino masses~\cite{Li:2010xr,Balazs:2010ha,Cheng:2012np}. Earlier GmSUGRA studies have considered electroweak supersymmetry, right-handed slepton bulk annihilation, Higgs/$Z$ funnels and light-neutralino collider prospects~\cite{Khan:2023ryc,Khan:2025azf,Khan:2025ibo,Khan:2026ewsusy}. The present analysis focuses instead on the heavier bino--wino branch associated with the LHC soft-lepton mass scale. The neutralino composition is determined from the mixing matrix, and the Planck-compatible and underabundant solutions are kept separate because their freeze-out mechanisms and collider cascades are not the same.

The scan yields two closely related classes. Planck-compatible points contain a bino LSP at the $99.8$--$99.95\%$ level and a wino $\widetilde\chi_2^0/\widetilde\chi_1^\pm$ pair at the $99.7$--$99.93\%$ level, with $\dmchi$ concentrated near $19$~GeV. Underabundant solutions retain the same electroweakino composition but contain a stau within about $6$--$8\%$ of the LSP, introducing an additional coannihilation channel and opening stau-mediated electroweakino decays. These spectra are compared with the ATLAS and CMS compressed-search results and with direct-detection constraints without replacing the experimental likelihood by a mass-plane overlay.

The paper is organized as follows. Section~\ref{sec:model} describes the GmSUGRA boundary conditions and the neutralino composition. Section~\ref{sec:scan} presents the numerical scan and the phenomenological requirements. Section~\ref{sec:collider} discusses the compressed spectrum in relation to the LHC soft-lepton searches. Sections~\ref{sec:red} and~\ref{sec:green} examine the Planck-compatible and underabundant solutions. Direct detection and comparison with recent collider studies are discussed in Secs.~\ref{sec:dd} and~\ref{sec:discussion}. Section~\ref{sec:conclusion} summarizes the results.

\section{GmSUGRA framework and compressed electroweakinos}
\label{sec:model}

GmSUGRA relaxes universal soft supersymmetry-breaking parameters while retaining GUT-scale relations among gauge couplings and gaugino masses~\cite{Li:2010xr,Balazs:2010ha}. At the unification scale one may write
\begin{align}
 \frac{1}{\alpha_2}-\frac{1}{\alpha_3}
 &=k\left(\frac{1}{\alpha_1}-\frac{1}{\alpha_3}\right),\\
 \frac{M_2}{\alpha_2}-\frac{M_3}{\alpha_3}
 &=k\left(\frac{M_1}{\alpha_1}-\frac{M_3}{\alpha_3}\right).
\label{eq:gmsugra}
\end{align}
For $k=5/3$ and unified gauge couplings, the gaugino relation becomes
\begin{equation}
 M_2-M_3=\frac53(M_1-M_3),
 \qquad
 M_3=\frac52 M_1-\frac32 M_2.
\label{eq:gaugino}
\end{equation}
Two of the three gaugino masses are therefore independent. The relation permits $M_1$ and $M_2$ to evolve to nearby bino and wino masses at the electroweak scale while the corresponding gluino remains in the multi-TeV range.

The scalar sector is nonuniversal as well. In the $SU(5)$ realization used here, the squark soft masses satisfy~\cite{Balazs:2010ha,Cheng:2012np}
\begin{align}
 m_{\widetilde Q_i}^2&=\frac56(m_0^U)^2+\frac16m_{\widetilde E_i^c}^2,\\
 m_{\widetilde U_i^c}^2&=\frac53(m_0^U)^2-\frac23m_{\widetilde E_i^c}^2,\\
 m_{\widetilde D_i^c}^2&=\frac53(m_0^U)^2-\frac23m_{\widetilde L_i}^2.
\label{eq:scalar}
\end{align}
The slepton masses, the two Higgs soft masses and the trilinear terms are taken as independent inputs. The parameter set is
\begin{equation}
\{m_0^U,m_{\widetilde E^c},m_{\widetilde L},M_1,M_2,
 A_t=A_b,A_\tau,m_{H_d},m_{H_u},\tan\beta\},
\label{eq:inputs}
\end{equation}
with $M_3$ obtained from Eq.~\eqref{eq:gaugino}. The two signs of $\mu$ are treated separately.

At the electroweak scale the neutralino mass matrix in the basis $(\widetilde B,\widetilde W^0,\widetilde H_d^0,\widetilde H_u^0)$ is
{\small
\begin{equation}
{\cal M}_N=
\setlength{\arraycolsep}{1.6pt}\begin{pmatrix}
M_1&0&-m_Zs_Wc_\beta&m_Zs_Ws_\beta\\
0&M_2&m_Zc_Wc_\beta&-m_Zc_Ws_\beta\\
-m_Zs_Wc_\beta&m_Zc_Wc_\beta&0&-\mu\\
m_Zs_Ws_\beta&-m_Zc_Ws_\beta&-\mu&0
\end{pmatrix}.
\label{eq:neutralino_matrix}
\end{equation}}
For the benchmark spectra $|M_1|\lesssim|M_2|\ll|\mu|$ after renormalization-group evolution. The two lightest neutralinos are consequently close to the bino and neutral-wino gauge eigenstates, whereas the Higgsino-like neutralinos are multi-TeV.

The composition is obtained from the neutralino mixing matrix rather than inferred from the mass ordering. In the conventional notation, the eight benchmarks satisfy
\begin{align}
 |N_{11}|^2&=0.9983\text{--}0.9995,\\
 |N_{22}|^2&=0.9973\text{--}0.9993,
\label{eq:fractions}
\end{align}
where $|N_{11}|^2$ is the bino probability of $\widetilde\chi_1^0$ and $|N_{22}|^2$ is the wino probability of $\widetilde\chi_2^0$. The lightest chargino is wino like and nearly degenerate with $\widetilde\chi_2^0$. This hierarchy fixes both the gauge strength of $\widetilde\chi_2^0\widetilde\chi_1^\pm$ production and the thermal role of the nearby wino states.

\section{Numerical scan and phenomenological requirements}
\label{sec:scan}

The supersymmetric spectrum is evaluated with ISAJET~7.85~\cite{Baer:1999sp,Paige:2003mg}. Gauge and third-generation Yukawa couplings are evolved from the electroweak scale to the unification scale, defined by $g_1=g_2$. A small mismatch with $g_3$ is allowed to account for GUT-scale threshold effects~\cite{Hisano:1992jj,Yamada:1992kv,Allanach:2006fy}. The GmSUGRA boundary conditions are applied to the soft terms and the two-loop MSSM renormalization-group equations are evolved back to the electroweak scale. The iteration includes the supersymmetric threshold corrections and is continued until a consistent spectrum is obtained. Radiative electroweak symmetry breaking is imposed and the lightest supersymmetric particle is required to be a neutralino.

The random scan covers
\begin{align}
100~{\rm GeV}&\le m_0^U\le5~{\rm TeV},\nonumber\\
100~{\rm GeV}&\le M_1\le700~{\rm GeV},\nonumber\\
100~{\rm GeV}&\le M_2\le900~{\rm GeV},\nonumber\\
100~{\rm GeV}&\le m_{\widetilde L},m_{\widetilde E^c}\le1.5~{\rm TeV},\nonumber\\
0&\le m_{H_u},m_{H_d}\le10~{\rm TeV},\nonumber\\
-16~{\rm TeV}&\le A_t=A_b\le16~{\rm TeV},\nonumber\\
-6~{\rm TeV}&\le A_\tau\le6~{\rm TeV},\qquad 2\le\tan\beta\le60.
\label{eq:scanranges}
\end{align}
Both signs of $\mu$ are scanned. Metropolis--Hastings sampling is used over the broad domain, followed by denser sampling around the compressed bino--wino solutions. The top-quark pole mass is fixed at $m_t=173.3$~GeV. This procedure follows the GmSUGRA spectrum treatment used in Refs.~\cite{Khan:2025ibo,Muhammad:2026ps}.

The light CP-even Higgs mass is restricted to
\begin{equation}
122~{\rm GeV}\leq m_h\leq128~{\rm GeV},
\label{eq:higgsbound}
\end{equation}
which allows for the theoretical uncertainty of the MSSM Higgs calculation~\cite{Slavich:2020zjv}. Charged sparticle masses are required to satisfy the LEP2 lower limits, and the colored spectrum is subjected to the working requirements
\begin{equation}
\begin{aligned}
 m_{\widetilde g}&\gtrsim2.2~{\rm TeV}, &
 m_{\widetilde t_1}&\gtrsim1.25~{\rm TeV},\\
 m_{\widetilde b_1}&\gtrsim1.5~{\rm TeV}, &
 m_{\widetilde q}&\gtrsim2.0~{\rm TeV},
\end{aligned}
\label{eq:coloredbounds}
\end{equation}
consistent with the collider selection used for the scan and with the corresponding Run-2 searches~\cite{ATLAS:2020dsf,ATLAS:2019gdh,Khan:2025ibo,Muhammad:2026ps}. The electroweak spectrum is compared with the relevant slepton and chargino/neutralino searches~\cite{ATLAS:2019slepton,ATLAS:2020compressed,ATLAS:2021threelep,CMS:2022soft,CMS:2024combined}.

The flavor requirements are
\begin{align}
0.8\times10^{-9}&\leq {\rm BR}(B_s\to\mu^+\mu^-)\leq6.2\times10^{-9},\\
2.99\times10^{-4}&\leq {\rm BR}(b\to s\gamma)\leq3.87\times10^{-4},\\
0.15&\leq\frac{{\rm BR}(B_u\to\tau\nu)_{\rm MSSM}}
{{\rm BR}(B_u\to\tau\nu)_{\rm SM}}\leq2.41,
\label{eq:flavor}
\end{align}
which are the working two- or three-standard-deviation intervals used in the numerical selection~\cite{LHCb:2012skj,HFLAV:2012imy,HFLAV:2010pgm,Muhammad:2026ps, Khan:2025zps,Ahmed:2022ude}.

For the neutralino relic abundance we retain the calculated value $\Omegachi$ and separate two samples. The Planck-compatible set satisfies
\begin{equation}
0.114\leq\Omegachi\leq0.126,
\label{eq:planck}
\end{equation}
corresponding to the working $5\sigma$ interval around the Planck 2018 determination used in the scan~\cite{Planck:2018vyg,Khan:2025ibo,Muhammad:2026ps}. A second set satisfies all non-cosmological requirements but has
\begin{equation}
\Omegachi<0.114.
\label{eq:under}
\end{equation}
These points are referred to as underabundant. Spin-independent (SI) and spin-dependent (SD) neutralino--proton cross sections are compared with the XENONnT and LZ results and the projected LZ exposure~\cite{XENON:2023cxc,LZ:2022lsv,LZ:2024zvo,LZ:2018qzl}.

\section{Compressed spectrum and the LHC soft-lepton channel}
\label{sec:collider}

\begin{figure*}[t]
\centering
\includegraphics[width=0.48\textwidth]{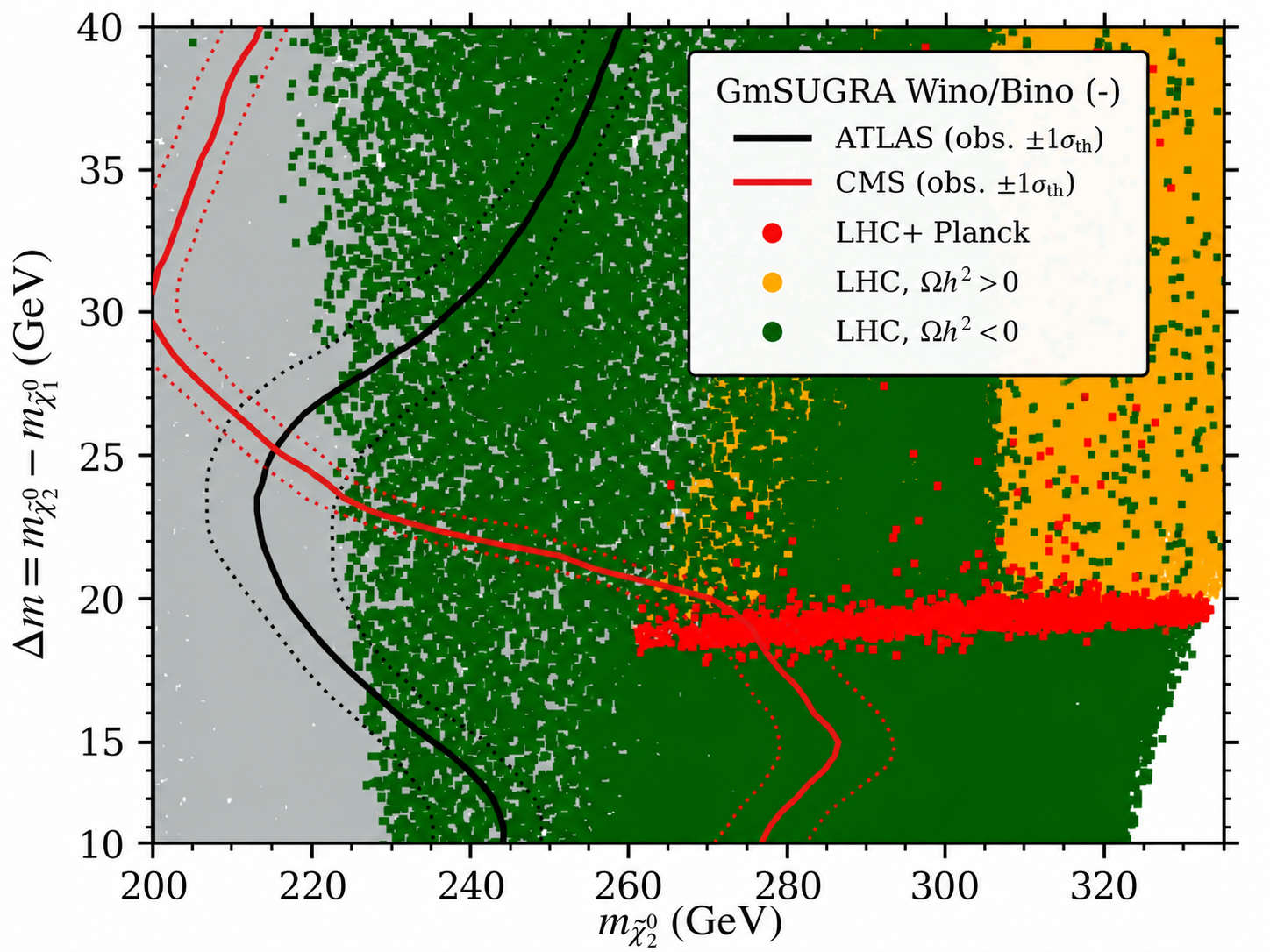}\hfill
\includegraphics[width=0.48\textwidth]{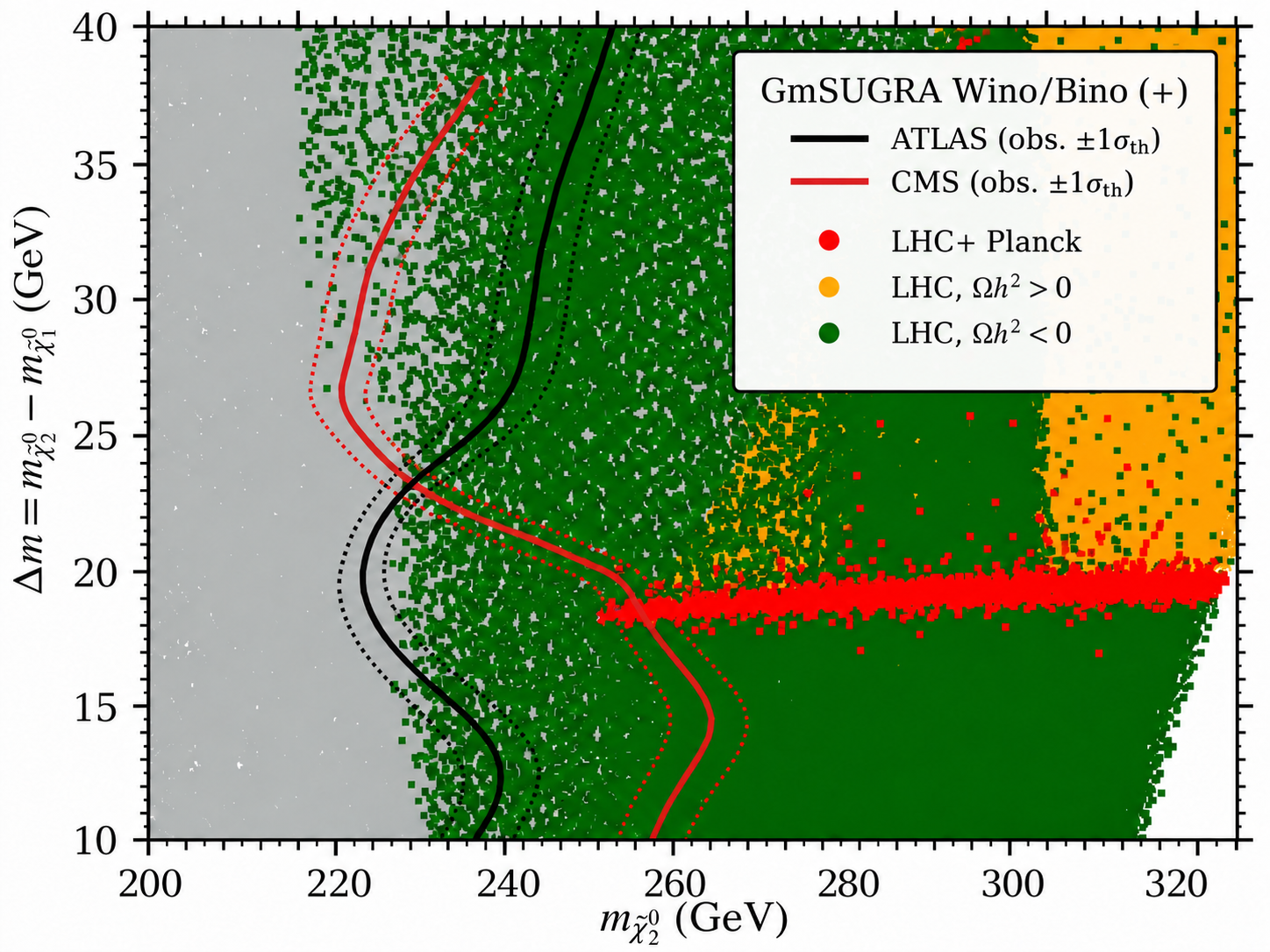}
\caption{$\mchiTwo$ versus $\dmchi=\mchiTwo-\mchi$ for the $\mu<0$ (left) and $\mu>0$ (right) scans. Red points satisfy the collider, Higgs, and flavor requirements and the Planck interval in Eq.~\eqref{eq:planck}; green and orange points satisfy the same non-cosmological requirements with $\Omegachi<0.114$, and $\Omegachi>0.126$, respectively. The overlaid ATLAS and CMS curves are the observed wino-bino $(-,+)$ simplified-model contours discussed in Ref.~\cite{Chakraborti:2024pdn} and based on Refs.~\cite{ATLAS:2020compressed,ATLAS:2021threelep,CMS:2022soft}.}
\label{fig:compressed}
\end{figure*}

Figure~\ref{fig:compressed} shows the scan in the $(\mchiTwo,\dmchi)$ plane for $\mu<0$ and $\mu>0$. The red points satisfy the collider, Higgs and flavor requirements together with Eq.~\eqref{eq:planck}; the green points satisfy the same non-cosmological requirements with $\Omegachi<0.114$. The Planck-compatible population forms a narrow band around $\dmchi\simeq19$~GeV, whereas the underabundant sample extends to both smaller and larger splittings and includes the $22$--$25$~GeV benchmarks discussed below. The appearance of the red band follows from the weak-scale bino--wino separation generated by Eq.~\eqref{eq:gaugino}, rather than from fixing $\dmchi$ as an electroweak-scale input.

The observed and expected ATLAS/CMS contours shown in Fig.~\ref{fig:compressed} are those used in the wino--bino interpretation of Ref.~\cite{Chakraborti:2024pdn}, based on the corresponding soft-lepton searches~\cite{ATLAS:2020compressed,ATLAS:2021threelep,CMS:2022soft}. Their $(+)$ and $(-)$ labels refer to the relative electroweak-scale sign of $M_1M_2$. The present scan has $M_1M_2>0$ for both signs of $\mu$, so the wino--bino $(+)$ comparison is the sign-consistent one in both panels. The sign of $\mu$ remains relevant through Higgsino mixing and Higgs-mediated scattering, but it is not the sign convention used for the collider contour.

The experimental sensitivity follows directly from the decay kinematics. For the red benchmarks, $\mchiTwo=278.6$--$285.2$~GeV, $\mchip=279.5$--$285.9$~GeV and $\mchi=259.7$--$266.2$~GeV. The neutralino splitting is $18.5$--$19.1$~GeV and the chargino--LSP splitting is of the same order. Since both are below the on-shell $Z$ and $W$ thresholds, the gauge-boson decays populate the low-$p_T$ two- and three-lepton signal regions. The corresponding dilepton endpoint lies near $19$~GeV, within the invariant-mass interval emphasized by phenomenological studies of the Run-2 soft-lepton data~\cite{Agin:2024coherent,Chakraborti:2024pdn,Agin:2025joint}.

For these Planck-compatible spectra the sleptons are heavier than $\widetilde\chi_2^0$ and $\widetilde\chi_1^\pm$, so two-body slepton cascades are closed. The off-shell $Z/W$ topology is therefore the appropriate kinematic reference for the red branch. The exact branching fractions and signal efficiencies remain spectrum dependent and require a point-by-point collider recast. The wino purity of the produced states is nevertheless relevant: associated $\widetilde\chi_2^0\widetilde\chi_1^\pm$ production retains the electroweak gauge coupling, whereas the bino LSP has no comparable direct production channel.

The older soft-lepton contours remove a portion of the lower-mass compressed population and place the red benchmark region near the transition between excluded and unexcluded simplified-model parameter space. More recent Run-2 combinations treat a broader set of electroweakino final states and can be more restrictive than a single contour~\cite{CMS:2024combined,ATLAS:2024combination}. The 2025 CMS soft-lepton update extends electron reconstruction to $p_T$ near $1$~GeV and quotes its largest gain at splittings of a few GeV~\cite{CMS:2025softupdate}; the low-momentum lepton-track and isolated-track analyses probe still smaller splittings~\cite{CMS:2025leptrack,CMS:2026track}. Those searches complement rather than replace the two/three soft-lepton analyses in the $\sim20$~GeV region considered here.

\section{Planck-compatible bino-wino solutions}
\label{sec:red}

The four Planck-compatible benchmarks are listed in Table~\ref{tab:red}. All four contain a nearly pure bino LSP, a nearly pure wino $\widetilde\chi_2^0/\widetilde\chi_1^\pm$ pair and Higgsino-like states above about $1.8$~TeV. Although $\mu$, the heavy-Higgs masses and the slepton spectrum vary among the benchmarks, $\dmchi$ remains within $18.5$--$19.1$~GeV. The stability of this separation across the four points reflects the gaugino relation more directly than Higgsino mixing.

The thermal behavior follows the standard coannihilation formalism~\cite{Griest:1990kh,Edsjo:1997bg}. For a state $i$ close to the LSP, with $\Delta_i=(m_i-\mchi)/\mchi$, the equilibrium contribution is weighted approximately by
\begin{equation}
 r_i(x)\propto g_i(1+\Delta_i)^{3/2}e^{-x\Delta_i},
 \qquad x\equiv\frac{\mchi}{T}.
\label{eq:boltzmann}
\end{equation}
For the red benchmarks the wino mass gap corresponds to $\Delta_{\widetilde W}\simeq0.070$--$0.073$. At $x\sim20$--$25$, the Boltzmann factor remains at the level of roughly $0.17$--$0.25$. The wino-like neutralino and chargino therefore remain populated during freeze-out and their gauge interactions contribute to the effective annihilation rate. A larger separation suppresses these contributions and raises the bino abundance, whereas a smaller separation increases the coannihilation efficiency. The Planck band in Eq.~\eqref{eq:planck} selects the narrow mass difference realized by the red points. The Planck-compatible solution is therefore characterized primarily by bino-wino freeze-out. The same mass difference also fixes the visible energy released in the $Z^{(*)}/W^{(*)}$ decays. The relic-density condition and the soft-lepton kinematics consequently select the same spectral quantity, $\dmchi$.

\begin{table*}[!t]
\caption{Representative Planck-compatible GmSUGRA solutions. All masses and dimensionful inputs are in GeV unless indicated otherwise. $f_B$ and $f_W$ denote the bino fraction of $\widetilde\chi_1^0$ and the wino fraction of $\widetilde\chi_2^0$, respectively.}
\label{tab:red}
\centering
\footnotesize
\setlength{\tabcolsep}{3.8pt}
\renewcommand{\arraystretch}{0.90}
\begin{tabular}{lrrrr}
\toprule
Quantity & R-1 ($\mu<0$) & R-2 ($\mu<0$) & R+1 ($\mu>0$) & R+2 ($\mu>0$)\\
\midrule
\multicolumn{5}{c}{GUT-scale input parameters}\\
\midrule
$m_0^U$ & 4815.0 & 3558.0 & 4763.0 & 4780.0\\
$m_{\tilde E^c}$ & 1262.0 & 956.60 & 1076.0 & 1333.0\\
$m_{\tilde L}$ & 1376.0 & 1140.0 & 1473.0 & 1493.0\\
$M_1$ & 591.10 & 597.20 & 590.40 & 602.70\\
$M_2$ & 320.00 & 325.80 & 318.20 & 326.20\\
$M_3$ & 997.75 & 1004.3 & 998.70 & 1017.4\\
$A_t=A_b$ & -9577.0 & -8373.0 & -9220.0 & -9309.0\\
$A_\tau$ & -1776.0 & 4547.0 & 2504.0 & -1718.0\\
$m_{H_d}$ & 3931.0 & 797.80 & 3249.0 & 3398.0\\
$m_{H_u}$ & 6274.0 & 384.80 & 4934.0 & 5574.0\\
$\tan\beta$ & 34.400 & 25.300 & 44.700 & 38.100\\
\midrule
\multicolumn{5}{c}{Higgs and electroweakino sector}\\
\midrule
$\mu(Q)$ & -1875.5 & -4511.5 & 3474.5 & 2867.4\\
$m_h$ & 125.55 & 125.19 & 125.53 & 125.53\\
$m_A$ & 2172.2 & 3640.4 & 1948.1 & 2054.4\\
$m_{\tilde\chi_1^0}$ & 260.17 & 261.05 & 259.72 & 266.15\\
$m_{\tilde\chi_2^0}$ & 279.23 & 279.56 & 278.55 & 285.19\\
$m_{\tilde\chi_1^\pm}$ & 279.53 & 280.92 & 279.53 & 285.87\\
$m_{\tilde\chi_{3,4}^0}$ & 1854.8/1855.0 & 4477.9/4478.2 & 3422.8/3423.1 & 2859.9/2860.5\\
$\dmchi$ & 19.060 & 18.510 & 18.830 & 19.040\\
$f_B(\tilde\chi_1^0)$ [\%] & 99.834 & 99.832 & 99.880 & 99.872\\
$f_W(\tilde\chi_2^0)$ [\%] & 99.726 & 99.814 & 99.850 & 99.826\\
\midrule
\multicolumn{5}{c}{Colored and slepton spectrum}\\
\midrule
$m_{\tilde g}$ & 2441.0 & 2411.4 & 2447.4 & 2486.0\\
$m_{\tilde t_1}/m_{\tilde t_2}$ & 1944.2/3294.2 & 2218.2/3032.4 & 2048.8/3724.2 & 2120.4/3538.7\\
$m_{\tilde b_1}/m_{\tilde b_2}$ & 2048.4/5749.3 & 2271.6/4508.6 & 2132.0/5458.1 & 2212.1/5643.6\\
$m_{\tilde u_L}/m_{\tilde u_R}$ & 4771.8/6327.5 & 3775.8/4796.2 & 4735.3/6251.3 & 4763.6/6274.9\\
$m_{\tilde d_L}/m_{\tilde d_R}$ & 4772.5/6525.4 & 3776.7/5010.7 & 4735.9/6478.6 & 4764.3/6495.7\\
$m_{\tilde e_L}/m_{\tilde e_R}$ & 1026.5/1730.0 & 729.31/1549.1 & 1003.5/1791.1 & 1150.7/1826.4\\
$m_{\tilde\mu_L}/m_{\tilde\mu_R}$ & 1026.5/1730.0 & 729.31/1549.1 & 1003.5/1791.1 & 1150.7/1826.4\\
$m_{\tilde\tau_1}/m_{\tilde\tau_2}$ & 784.60/1422.7 & 345.09/1254.0 & 477.27/1263.3 & 933.66/1540.3\\
\midrule
\multicolumn{5}{c}{Dark-matter observables}\\
\midrule
$\sigma_p^{\rm SI}$ [pb] & $3.2332\times10^{-14}$ & $2.0445\times10^{-13}$ & $1.4658\times10^{-13}$ & $3.0716\times10^{-13}$\\
$\sigma_p^{\rm SD}$ [pb] & $1.4062\times10^{-8}$ & $7.4216\times10^{-11}$ & $8.5434\times10^{-10}$ & $2.1523\times10^{-9}$\\
$\Omega_\chi h^2$ & 0.11957 & 0.12121 & 0.12147 & 0.11911\\
\bottomrule
\end{tabular}
\end{table*}

\section{Underabundant solutions and the stau-assisted branch}
\label{sec:green}

The underabundant benchmarks in Table~\ref{tab:green} retain the same neutralino identity as the Planck-compatible points. The LSP remains $99.9\%$ bino and $\widetilde\chi_2^0$ remains $99.8\%$ wino to the quoted precision, while the Higgsino-like neutralinos are in the multi-TeV range. Their smaller $\Omegachi$ is therefore not produced by a transition to a Higgsino-dominated LSP. The bino--wino separation is also slightly larger than in the red sample, with $\dmchi=22$--$25$~GeV, so the lower relic density cannot be assigned to a tighter bino--wino degeneracy.

The relevant additional state is $\widetilde\tau_1$. For the four green benchmarks,
\begin{equation}
\frac{m_{\widetilde\tau_1}-\mchi}{\mchi}=0.063\text{--}0.082.
\label{eq:staugap}
\end{equation}
At $x\sim20$--$25$ the stau population is only moderately Boltzmann suppressed. Stau--neutralino and stau-pair processes therefore supplement the wino-mediated channels in the effective freeze-out rate. This spectrum accounts for the lower values $\Omegachi=0.054$--$0.106$ without invoking a sizeable Higgsino component. A channel-by-channel decomposition is not assigned here because the stored scan observables do not contain individual annihilation fractions.

The stau ordering also changes the collider interpretation. In all four green benchmarks,
\begin{equation}
\mchi<m_{\widetilde\tau_1}<\mchiTwo\simeq\mchip.
\label{eq:greenordering}
\end{equation}
The two-body modes $\widetilde\chi_2^0\to\widetilde\tau_1\tau$ and $\widetilde\chi_1^\pm\to\widetilde\tau_1\nu_\tau$ are therefore kinematically open. Their branching fractions depend on the stau mixing and on the electroweakino couplings, but the mass ordering alone shows that the green branch need not follow the pure $Z^{(*)}W^{(*)}$ simplified topology. A collider test of these points requires the soft-lepton searches to be considered together with final states containing soft taus and light sleptons.

\begin{table*}[!t]
\caption{Representative underabundant GmSUGRA solutions satisfying $\Omega_\chi h^2<0.114$. The final row in the slepton block gives the stau--LSP fractional mass difference. All masses and dimensionful inputs are in GeV unless indicated otherwise. $f_B$ and $f_W$ denote the bino fraction of $\widetilde\chi_1^0$ and the wino fraction of $\widetilde\chi_2^0$, respectively.}
\label{tab:green}
\centering
\footnotesize
\setlength{\tabcolsep}{3.8pt}
\renewcommand{\arraystretch}{0.90}
\begin{tabular}{lrrrr}
\toprule
Quantity & G-1 ($\mu<0$) & G-2 ($\mu<0$) & G+1 ($\mu>0$) & G+2 ($\mu>0$)\\
\midrule
\multicolumn{5}{c}{GUT-scale input parameters}\\
\midrule
$m_0^U$ & 4860.0 & 4836.0 & 4736.0 & 4981.0\\
$m_{\tilde E^c}$ & 1381.0 & 921.00 & 927.60 & 1260.0\\
$m_{\tilde L}$ & 1484.0 & 1466.0 & 1402.0 & 1377.0\\
$M_1$ & 566.80 & 609.50 & 589.20 & 574.90\\
$M_2$ & 310.90 & 335.20 & 323.60 & 318.70\\
$M_3$ & 950.65 & 1020.9 & 987.60 & 959.20\\
$A_t=A_b$ & -9507.0 & -9226.0 & -8941.0 & -9264.0\\
$A_\tau$ & 1501.0 & 3975.0 & 3352.0 & -989.50\\
$m_{H_d}$ & 5050.0 & 4028.0 & 3450.0 & 3557.0\\
$m_{H_u}$ & 6005.0 & 6216.0 & 5526.0 & 4471.0\\
$\tan\beta$ & 38.200 & 36.300 & 37.500 & 40.000\\
\midrule
\multicolumn{5}{c}{Higgs and electroweakino sector}\\
\midrule
$\mu(Q)$ & -2242.8 & -1824.3 & 2714.9 & 4087.3\\
$m_h$ & 125.69 & 125.37 & 125.26 & 125.20\\
$m_A$ & 3330.9 & 1784.7 & 1969.6 & 3608.2\\
$m_{\tilde\chi_1^0}$ & 249.51 & 266.66 & 258.04 & 254.54\\
$m_{\tilde\chi_2^0}$ & 272.64 & 289.96 & 280.08 & 279.69\\
$m_{\tilde\chi_1^\pm}$ & 272.95 & 290.24 & 280.68 & 280.80\\
$m_{\tilde\chi_{3,4}^0}$ & 2215.0/2215.3 & 1802.9/1803.1 & 2710.0/2710.7 & 4029.1/4029.3\\
$\dmchi$ & 23.130 & 23.300 & 22.040 & 25.150\\
$f_B(\tilde\chi_1^0)$ [\%] & 99.880 & 99.866 & 99.900 & 99.948\\
$f_W(\tilde\chi_2^0)$ [\%] & 99.806 & 99.750 & 99.850 & 99.926\\
\midrule
\multicolumn{5}{c}{Colored and slepton spectrum}\\
\midrule
$m_{\tilde g}$ & 2342.8 & 2489.0 & 2420.9 & 2380.5\\
$m_{\tilde t_1}/m_{\tilde t_2}$ & 1746.3/3398.7 & 1963.0/3455.3 & 2125.2/3584.8 & 2541.7/4065.8\\
$m_{\tilde b_1}/m_{\tilde b_2}$ & 1866.9/5543.4 & 2058.6/5727.1 & 2213.8/5617.8 & 2622.9/5906.9\\
$m_{\tilde u_L}/m_{\tilde u_R}$ & 4793.8/6310.2 & 4798.7/6382.9 & 4694.5/6236.5 & 4898.0/6464.8\\
$m_{\tilde d_L}/m_{\tilde d_R}$ & 4794.5/6575.5 & 4799.4/6572.9 & 4695.2/6433.8 & 4898.7/6735.6\\
$m_{\tilde e_L}/m_{\tilde e_R}$ & 1010.1/1987.1 & 1047.4/1635.6 & 947.40/1649.6 & 708.24/2001.4\\
$m_{\tilde\mu_L}/m_{\tilde\mu_R}$ & 1010.1/1987.1 & 1047.4/1635.6 & 947.40/1649.6 & 708.24/2001.4\\
$m_{\tilde\tau_1}/m_{\tilde\tau_2}$ & 266.34/1396.1 & 288.42/747.64 & 274.34/989.70 & 271.71/1733.6\\
$(m_{\tilde\tau_1}-m_{\tilde\chi_1^0})/m_{\tilde\chi_1^0}$ [\%] & 6.745 & 8.160 & 6.317 & 6.746\\
\midrule
\multicolumn{5}{c}{Dark-matter observables}\\
\midrule
$\sigma_p^{\rm SI}$ [pb] & $3.8968\times10^{-14}$ & $7.0125\times10^{-14}$ & $3.8603\times10^{-13}$ & $5.5526\times10^{-15}$\\
$\sigma_p^{\rm SD}$ [pb] & $6.3381\times10^{-9}$ & $1.5222\times10^{-8}$ & $2.6291\times10^{-9}$ & $3.1369\times10^{-10}$\\
$\Omega_\chi h^2$ & 0.10218 & 0.10597 & 0.054503 & 0.062884\\
\bottomrule
\end{tabular}
\end{table*}

\section{Direct detection}
\label{sec:dd}

\begin{figure*}[th!]
	\centering \includegraphics[width=8.90cm]{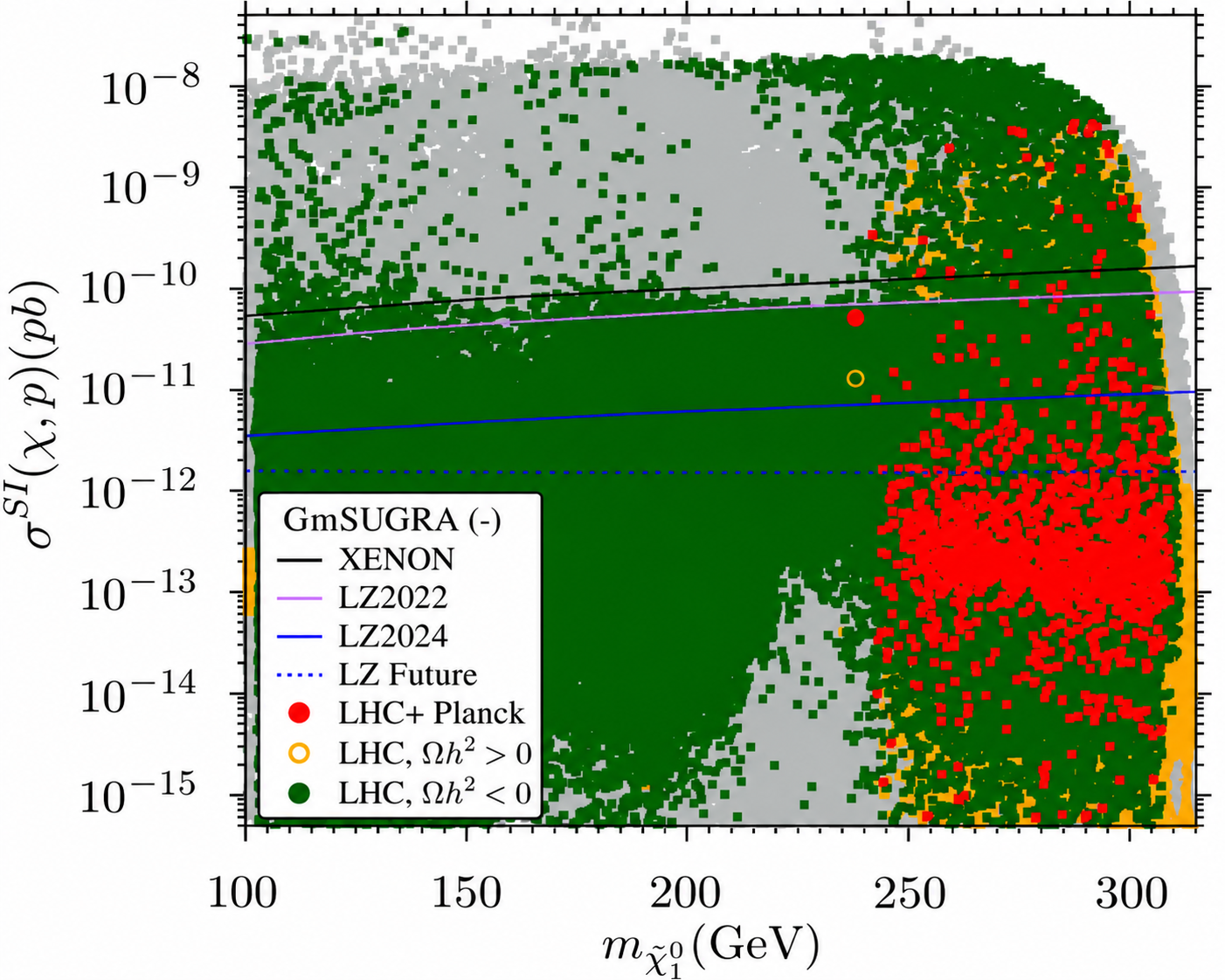}
    \centering \includegraphics[width=8.90cm]{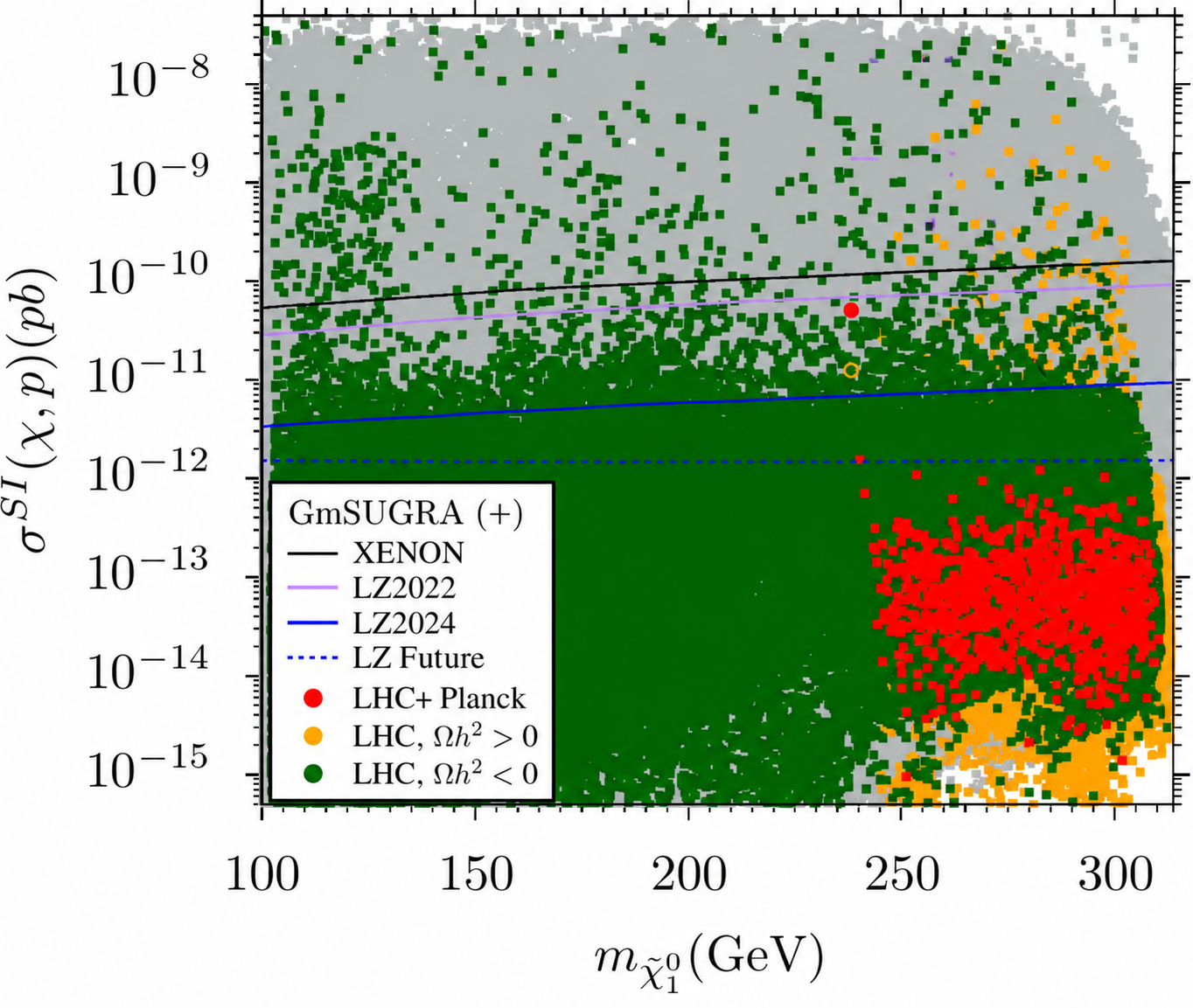}
	\centering \includegraphics[width=8.90cm]{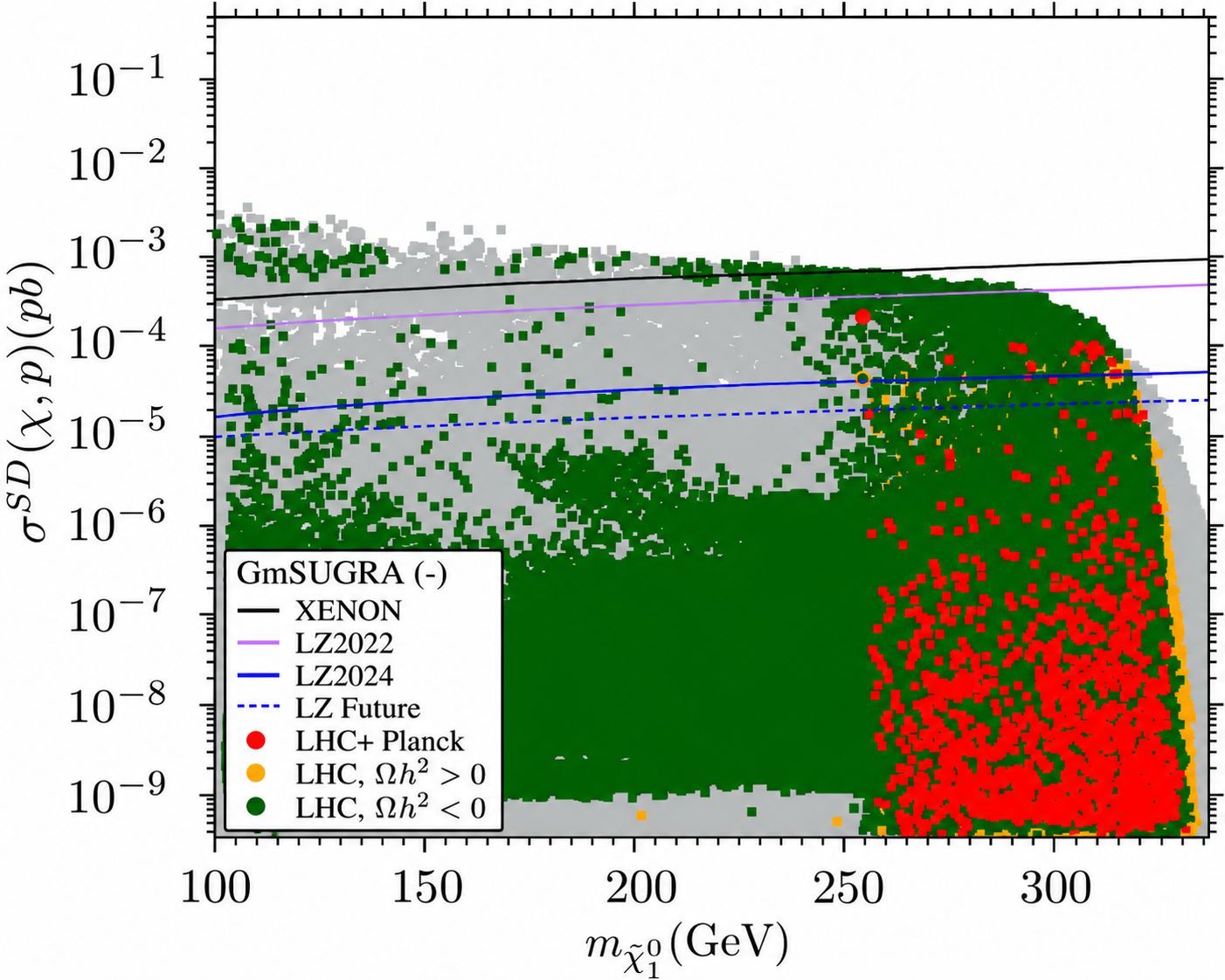}
    \centering \includegraphics[width=8.90cm]{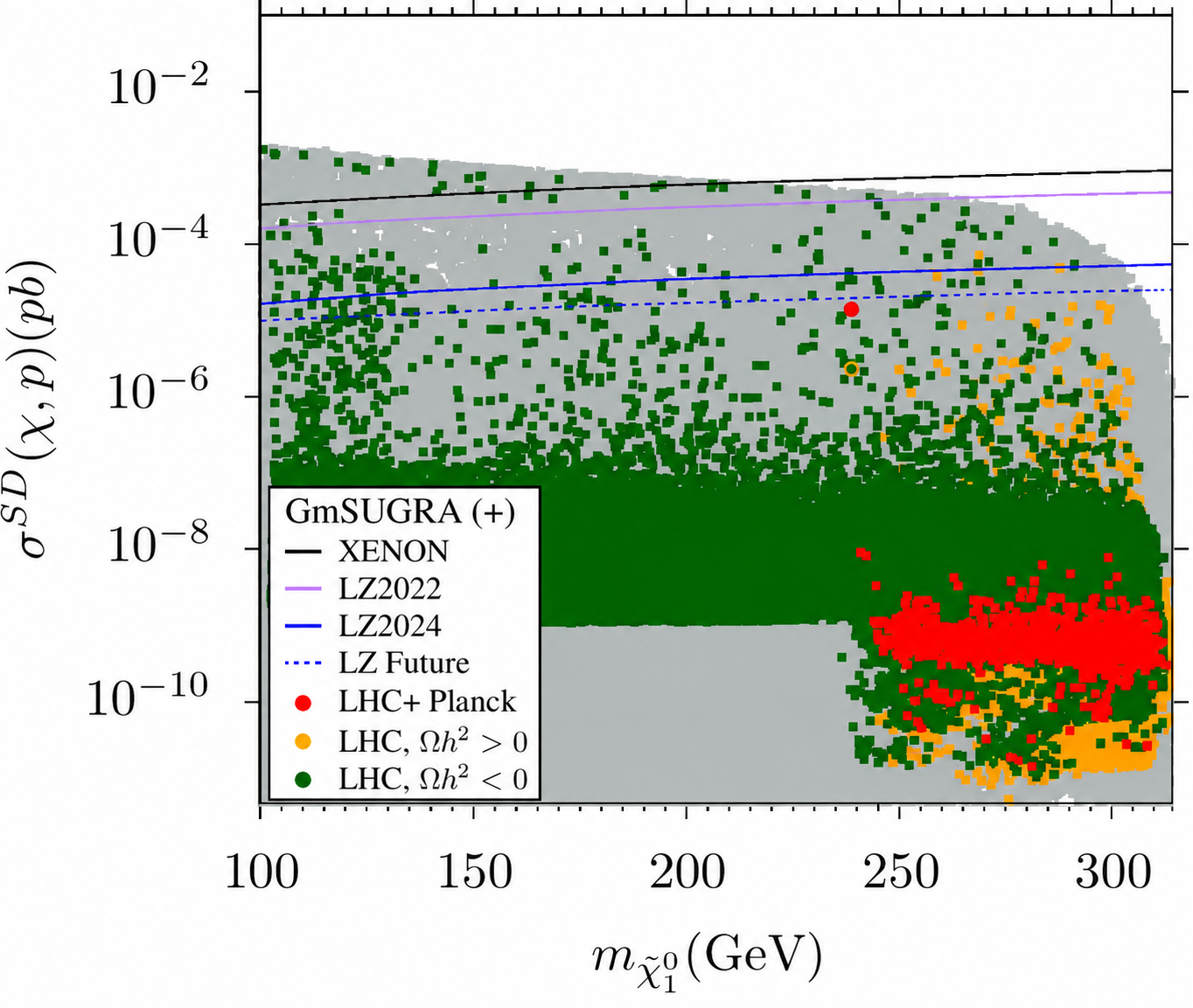}
    \caption{Neutralino--proton scattering cross sections as functions of $\mchi$. The upper panels show the SI results for $\mu<0$ (left) and $\mu>0$ (right), and the lower panels show the corresponding SD results. The curves denote the XENONnT result, the initial and 4.2-tonne-year LZ results, and the projected LZ exposure~\cite{XENON:2023cxc,LZ:2022lsv,LZ:2024zvo,LZ:2018qzl}. The underabundant points are plotted with the unrescaled particle-physics cross sections.}
\label{fig:dd}
\end{figure*}

Figure~\ref{fig:dd} shows the neutralino--proton SI and SD scattering cross sections. The LSP is nearly pure bino, so the SI amplitude arises through its small Higgsino admixture and the CP-even Higgs exchange. The scattering rate is consequently sensitive to $\mu$, $\tan\beta$ and the heavy-Higgs sector even when the bino and wino masses are similar. The two signs of $\mu$ therefore generate different interference patterns in the Higgs-mediated amplitude, consistent with the sign dependence found in light-neutralino GmSUGRA analyses~\cite{Khan:2025azf}.

For the Planck-compatible benchmarks, $\sigma_p^{\rm SI}$ ranges from $3.2\times10^{-14}$ to $3.1\times10^{-13}$~pb; the selected underabundant points span $5.6\times10^{-15}$--$3.9\times10^{-13}$~pb. The SD predictions cover a wider interval. The direct-detection observables thus probe a different combination of neutralino mixing parameters from the LHC production process, which is governed mainly by the wino content and the mass of $\widetilde\chi_2^0/\widetilde\chi_1^\pm$. The underabundant points are shown with the unrescaled cross sections returned by the spectrum calculation.

\section{Comparison with current collider studies}
\label{sec:discussion}

The Planck-compatible GmSUGRA benchmarks can be compared with the electroweak-scale wino--bino study of Ref.~\cite{Chakraborti:2024pdn}. That analysis places the soft-lepton region for the $M_1M_2>0$ configuration at $m_{\widetilde\chi_2^0}$ above roughly $250$~GeV with splittings of about $18$--$25$~GeV. The red GmSUGRA points, with $m_{\widetilde\chi_2^0}=279$--$285$~GeV and $\dmchi=18.5$--$19.1$~GeV, occupy the same kinematic interval. Ref.~\cite{Agin:2024coherent} associates several Run-2 soft-lepton fluctuations with dilepton masses around $10$--$20$~GeV, and the combined recast of Ref.~\cite{Agin:2025joint} selects a related wino--bino mass difference. In the GmSUGRA scan the $\sim19$~GeV separation is selected simultaneously by the high-scale gaugino relation and the relic-density requirement.

A different high-scale realization is obtained in the NMSSM analysis of Ref.~\cite{Bagnaschi:2026gut}. There the LSP is singlino dominated and the nearby neutralinos and chargino are Higgsino like, with an approximate conventional GUT relation among the gaugino masses. The present spectra instead contain a bino LSP, a wino $\widetilde\chi_2^0/\widetilde\chi_1^\pm$ pair and multi-TeV Higgsinos. Similar values of $\dmchi$ therefore correspond to different production couplings, relic-density mechanisms and direct-detection amplitudes.

The 2024 CMS and ATLAS combinations provide a broader Run-2 test than the individual soft-lepton contours~\cite{CMS:2024combined,ATLAS:2024combination}. Their published interpretations are still based on specified simplified spectra, whereas a GmSUGRA point contains correlated slepton, Higgs and electroweakino masses. This distinction becomes more pronounced for the green points because the stau is lighter than the wino-like states and opens additional two-body cascades. A detector-level recast must therefore use the spectrum-specific branching fractions rather than assign the simplified-model acceptance directly.

A recent global analysis of the MSSM electroweakino sector combines 34 Run-2 searches and 63 Run-2 measurements and reports $m_{\widetilde\chi_2^0}\gtrsim760$~GeV for a light bino LSP within an EWMSSM setup where the sfermions and non-SM Higgs states are decoupled~\cite{Athron:2026ewino}. It also reports that the separate compressed-spectrum fluctuations are not described simultaneously by the models considered. This result does not map one-to-one onto the present GmSUGRA spectra, especially the stau-assisted branch, but it shows that a single mass-plane contour is insufficient to establish collider viability. The red and green benchmarks are therefore formulated as spectra for a dedicated recast rather than as a claim of survival against the complete Run-2 likelihood.

The low-momentum CMS searches probe a complementary part of the compressed spectrum. The 2025 soft-lepton update extends sensitivity at $\Delta m$ of a few GeV~\cite{CMS:2025softupdate}; the lepton-track search addresses Higgsino splittings in the approximately $1$--$10$~GeV range~\cite{CMS:2025leptrack}; and the isolated-track analysis targets sub-GeV chargino--neutralino splittings~\cite{CMS:2026track}. The $18$--$25$~GeV GmSUGRA region remains associated with the two- and three-soft-lepton topology rather than with the near-degenerate track signatures. Run-3 analyses that retain low lepton thresholds while extending the integrated luminosity are therefore the relevant experimental continuation for this mass difference.

\section{Conclusions}
\label{sec:conclusion}

We have examined compressed bino--wino spectra generated from GmSUGRA boundary conditions for both signs of $\mu$. The selected weak-scale states are highly separated in composition: $\widetilde\chi_1^0$ is $99.8$--$99.95\%$ bino, $\widetilde\chi_2^0$ is $99.7$--$99.93\%$ wino, the lightest chargino is nearly degenerate with the wino-like neutralino, and the Higgsino-like states are multi-TeV.

The Planck-compatible solutions have $\mchi=260$--$266$~GeV, $\mchiTwo=279$--$285$~GeV and $\dmchi=18.5$--$19.1$~GeV. This separation is small enough for the wino states to remain thermally populated during freeze-out, so bino--wino coannihilation reduces the neutralino abundance into the interval $0.114\leq\Omegachi\leq0.126$. At the LHC the same separation places the gauge-boson decays below the on-shell $W/Z$ thresholds and gives a dilepton endpoint close to $19$~GeV. The cosmological and collider properties of the red branch are therefore controlled by the same bino--wino mass difference.

The underabundant solutions have $\dmchi=22$--$25$~GeV and $\Omegachi=0.054$--$0.106$. Their neutralino composition remains bino--wino, but a light stau lies only $6$--$8\%$ above the LSP. The resulting stau-assisted freeze-out lowers the relic abundance, while the ordering $\mchi<m_{\widetilde\tau_1}<\mchiTwo\simeq\mchip$ permits on-shell stau cascades. This branch therefore requires a collider treatment that includes soft-$\tau$ and slepton final states in addition to the standard $Z^{(*)}W^{(*)}$ topology.

The comparison with ATLAS and CMS places the red spectra in the mass-difference region addressed by the Run-2 soft-lepton searches and related phenomenological recasts. The Run-2 combinations and the recent global electroweakino analysis show that the quantitative status of a complete MSSM spectrum cannot be inferred from one simplified-model contour. A spectrum-specific recast, including production cross sections, branching fractions and the correlated signal-region likelihoods, is required for the final collider assessment. The benchmark sets supplied here define the GmSUGRA spectra for such an analysis and distinguish the bino--wino and stau-assisted thermal regimes.

\section{Acknowledgements}
TL is supported in part by the National Key Research and Development Program of China Grant No. 2020YFC2201504, by the Projects No. 11875062, No. 11947302, No. 12047503, and No. 12275333 supported by the National Natural Science Foundation of China, by the Key Research Program of the Chinese Academy of Sciences, Grant No. XDPB15, by the Scientific Instrument Developing Project of the Chinese Academy of Sciences, Grant No. YJKYYQ20190049, by the International Partnership Program of Chinese Academy of Sciences for Grand Challenges, Grant No. 112311KYSB20210012, and by the Henan Province Outstanding Foreign Scientist Studio Project, No.GZS2025008. This work was supported by the High Performance Computing Platform of Henan Normal University.
\bibliographystyle{apsrev4-2}
\bibliography{refs}

\end{document}